\documentclass[preprint,amsmath,amssymb]{article}
\usepackage{dcolumn}
\usepackage{bm}
\usepackage{hyperref}
\usepackage{amsmath} 
\usepackage{amsfonts}  
\usepackage{graphics}   
\usepackage{graphicx} 
\usepackage{epsfig}    
\usepackage{rotating}  
 \usepackage[latin1]{inputenc}
 \usepackage{color}
 \usepackage{rotating}
 \usepackage{setspace}
\usepackage{natbib}

\begin{document}

\begin{center}
{\Large \bf
Oscillatory Dynamics in Rock-Paper-Scissors Games with Mutations}\\
\end{center}

\vspace*{0.5cm}

\begin{center}
 {\large  Mauro Mobilia}\\

{\small 

Department of Applied Mathematics, School of Mathematics\\
University of Leeds\\
Leeds LS2 9JT, United Kingdom \\
Phone:+44-(0)11-3343-1591\\
Email: M.Mobilia@leeds.ac.uk\\
}
\end{center}

\newpage

{\bf Abstract}\\
We study the oscillatory dynamics in the generic three-species rock-paper-scissors  games with mutations. In the mean-field limit, different behaviors are found: (a) for high mutation rate, there is a stable interior fixed point with coexistence of all species; (b) for low mutation rates, there is a region of the parameter space characterized by a limit cycle resulting from a Hopf bifurcation; (c) in the absence of mutations, there is a region where heteroclinic cycles yield oscillations of large amplitude (not robust against  noise). After a discussion on the main properties of the mean-field dynamics, we investigate the stochastic version of the model within an individual-based formulation. Demographic fluctuations are therefore naturally accounted and their effects are studied using a diffusion theory complemented by numerical simulations. It is thus shown that persistent erratic oscillations (quasi-cycles) of large amplitude  emerge from a noise-induced resonance phenomenon. We also analytically and numerically compute the average escape time necessary to reach a (quasi-)cycle on which the system oscillates at a given amplitude.\\
\vspace*{0.3cm}


\newpage

\section{Introduction} 
Understanding the mechanisms allowing co-evolution and the maintenance of biodiversity is a key issue in theoretical biology and ecology~(\cite{May,Michod,Hubbell,Haken,Murray,Neal}). In this context, cyclic dominance  has been identified as a potential mechanism that helps promote species diversity~(\cite{durrett-1994-46,durrett-1997-185,gilg-2003-302,kerr-2002-418,czaran-2002-99})
and is naturally investigated in the framework of evolutionary game theory (EGT)~(\cite{Smith,Hofbauer,Nowak,szabo-2007-446}). The emergence of oscillatory (or quasi-oscillatory) behavior is one of the most appealing and debated phenomena that often characterizes co-evolution in population dynamics~(\cite{sinervo-1996-340,zamudio,kerr-2002-418,kirkup-2004-428,Dawkins,Bastolla,hauert-2002-296,Mobilia-2007}). Oscillatory dynamics has notably been  observed in predator-prey and host-pathogen systems~(\cite{Anderson1,Anderson2,Berryman,Turchin}), as well as in genetic networks~(\cite{Leibler}). Here, we study the oscillatory  dynamics of rock-paper-scissors  games with mutations and demonstrate that this favors the long-lasting co-evolution of all species. 
Rock-paper-scissors games (RPS) --in which rock crushes scissors, scissors cut paper, and paper wraps rock~(\cite{Hofbauer,Nowak,szabo-2007-446})-- have emerged as paradigmatic mathematical models in EGT to describe  the cyclic
competition in ecosystems~(\cite{tainaka-1994-50,may-1975-29,szabo-2002-65,reichenbach-2006-74,reichenbach-2007-448,reichenbach-2008,perc-2007-75,Claussen,Alava,Berr}). In addition to their theoretical relevance, and in spite of their apparent simplicity, it has been argued that the  RPS model and the like can  help understand the co-evolutionary dynamics
of different biological systems, such as the cyclic dominance observed in some communities of lizards~(\cite{sinervo-1996-340,zamudio}). Another popular example is the cyclic competition between three strains of {\it E.coli}~(\cite{kerr-2002-418}). In this case, it has been found that 
cyclic dominance yields  coexistence of all species in a spatial setting, while in a well-mixed (homogeneous) environment two species go extinct after a short transient. 
Naturally therefore, the mathematical properties of the RPS model and of its variants have recently attracted much interest. In an EGT setting, the RPS dynamics is classically described in terms of the replicator equations (REs) that is a set 
 of deterministic rate equations (see below)~(\cite{Hofbauer,Nowak,szabo-2007-446}). The latter essentially predict two types of behavior for the RPS dynamics: there is either the stable coexistence of all species, or  oscillations of large amplitudes with each species  almost taking over the entire system in turn and then suddenly almost going extinct. As population dynamics always involves a finite (yet, often large) number of discrete entities, it is necessary to take the effects of noise into account and understand its influence on the ensuing nonlinear dynamics~(\cite{Nisbet,Ewens}). In the context of population models, demographic stochasticity (i.e. intrinsic noise) is naturally accounted by adopting an `individual-based'  modeling. The 
stochastic dynamics is thus implemented in terms of random birth and death events, like in the Moran process~(\cite{moran-1958-54,Nowak}). Recently, individual-based modeling has been extensively used to study the stochastic RPS dynamics. In fact, a large body of  research has been concerned with spatially-extended systems and the influence of the species' movement on the emerging noisy patterns~(\cite{tainaka-1994-50,reichenbach-2007-448,reichenbach-2007-99,reichenbach-2008,Alava}). On the other hand,
for systems with homogeneous (well-mixed) populations, it is well established that intrinsic noise drastically alters the replicator dynamics and recent studies have focused on the mean time and probability of extinction of 
two species~(\cite{reichenbach-2006-74,Claussen,Berr}).

In addition to selection and reproduction, a third basic evolutionary mechanism is {\it mutation}. The latter is generally regarded as providing the advantageous traits that survive and multiply in offsprings~(\cite{Neal,Michod}). Also, from a behavioral perspective, it is recognized that agents never behave perfectly rationally and it is sensible to assume that individuals can switch their strategies and therefore undergo mutations~(\cite{Antal,szabo-2007-446}).
 While selection and reproduction are present in the RPS models considered in Refs.~(\cite{tainaka-1994-50,szabo-2002-65,reichenbach-2006-74,reichenbach-2007-448,perc-2007-75,Claussen,reichenbach-2007-99,reichenbach-2008,Alava,Berr}), the latter {\it do not} account for mutations. Here,  we study the oscillatory dynamics of RPS games with mutations and show that the possibility to switch  from one strategy (species) to another with a small transition rate yields novel oscillatory dynamics and favors long-term coexistence. For population of finite size, we discuss the effect of demographic noise on the dynamics and show that it induces
persistent (quasi-)cyclic behavior characterized by sustained erratic oscillations of non-vanishing amplitude.

The remainder of the paper is organized along the following lines. In the next section we introduce the generic RPS model with mutations (RPSM) and describe its dynamics in the mean-field limit. The  (generalized) REs are studied in Section 3, where the 
bifurcation diagram is obtained (Sec.~3.1) and a new region of the parameter space characterized by a Hopf bifurcation is identified. The
main properties of the resulting 
  limit cycle are briefly discussed in Sec.~3.2. Section 4 is dedicated to the stochastic dynamics of the model RPSM in terms of an individual-based formulation. In particular, we show that demographic noise can cause quasi-cyclic behavior with persistent oscillations of large amplitude (Sec.~4.1), and
allow to escape -- after a characteristic time that is computed -- from the coexistence fixed point and reach a given (quasi-)cycle (Sec.~4.2). In the final section, we summarize and discuss our results.

\section{Rock-Paper-Scissors with Mutations} 

In their essence, all variants of the RPS game aim at describing the co-evolutionary dynamics of three species, say $A, B$ and $C$, in cyclic competition. In this setting, as in the children's game, `rock' (here species $A$) crushes the `scissors' (species $B$), and  `paper' (species $C$) wraps the `rock', and `scissors' cut the `paper'. We therefore say that $A$ dominates over $C$, which outcompetes $B$, which outgrows $A$ and thus closes the cycle.
In EGT, the interactions are  specified in terms of a payoff matrix ${\cal P}$.
Generically, the  cyclic dominance of RPS games
is captured by the following payoff matrix~(\cite{Hofbauer,Nowak,szabo-2007-446}), where $\epsilon >0$: 
\\
\begin{table}[h]
\begin{center}
${\cal P}=$
\begin{tabular}{c|c  c  c }
vs & Rock $(A)$  & Paper $(B)$ & Scissors $(C)$ \\
 \hline
Rock $(A)$ & $0$ & $-\epsilon$ & $1$ \\ 
Paper  $(B)$ & $1$ & $0$ & $-\epsilon$ \\ 
Scissors $(C)$  & $-\epsilon$ & 1 & 0 \\ 
\end{tabular}
\end{center}
\end{table}
\\
According to this  matrix, when a pair of $A$ and $B$ players interacts,  the former gets a negative payoff $-\epsilon$ while the latter gets a payoff $1$. In this case, $A$ is dominated by $B$  and  its loss is less than $B$'s gain when $0<\epsilon<1$, whereas $B$'s gain is higher than  $A$'s loss  when $\epsilon>1$. In the same way, $C$'s dominate over $B$'s and the latter  prevail against $A$'s, and thus the model exhibits cyclic dominance. In all pairwise interactions, the dominant individual  gets a payoff $1$, whereas the dominated one obtain a negative payoff $-\epsilon$.
Therefore, the parameter $\epsilon$ allows to introduce an asymmetry (when $\epsilon\neq 1$) in the interactions. When $\epsilon=1$, one of the player loses what the other gains and 
this perfect balance corresponds to a zero-sum game. 

In EGT, one often considers homogeneous (well-mixed) populations of $N$ individuals, with $N\to \infty$. In this mean-field limit,
the dynamics is
usually specified
in terms of the REs for the densities (or relative abundances) $a(t), b(t)$ and $c(t)$ of species $A, B$ and $C$, respectively~(\cite{Hofbauer,Nowak,szabo-2007-446}). 
Introducing the vector ${\bm s}(t)$, whose components $s_i$, with $\;i \in (A,B,C)$, are $s_A\equiv a(t), s_B\equiv b(t)$ and $s_C\equiv c(t)$, the REs read
 $\dot{s_i}=s_i[ ({\cal P} {\bm s})_i - {\bm s}.{\cal P} {\bm s}]= s_i[ \pi_i - \bar{\pi}]$, where  the dot stands for 
the time derivative. The important notion of average payoff (per individual) of species $i$, $\pi_i$,  has been introduced in terms of the payoff matrix as a linear function of the relative abundances:
$\pi_i\equiv ({\cal P} {\bm s})_i$, whereas 
 $\bar{\pi}= {\bm s}.{\cal P} {\bm s}=\sum_{i} s_i \pi_i$
denotes the population's mean payoff~(\cite{Hofbauer,Nowak,szabo-2007-446}).
In addition to the above processes of selection/reproduction, 
we introduce a third evolutionary mechanism that allows each individual to mutate from one species to another  with rate $\mu$:
\begin{eqnarray}
\label{react}
A \stackrel{\mu}{\longrightarrow}
\begin{cases}
B\\
C
\end{cases}
\,, \quad
B \stackrel{\mu}{\longrightarrow}
\begin{cases}
A\\
C
\end{cases}
\,, \quad
C \stackrel{\mu}{\longrightarrow}
\begin{cases}
A\\
B
\end{cases}. 
\end{eqnarray}
In this setting, the natural generalization of the replicator equations for the model under consideration is 
$\dot{s}_i=s_i[ \pi_i - \bar{\pi}]+ \mu(1-3s_i),$ with $\pi_A=c  -\epsilon b,\;\pi_B=a  -\epsilon c, \;\pi_C=b  -\epsilon a$ and  $\bar{\pi}=(1-\epsilon)(ab + bc + ac)$. The latter are studied in the next sections.
\begin{figure}[h]   
\begin{center} 
\includegraphics[scale=0.267]{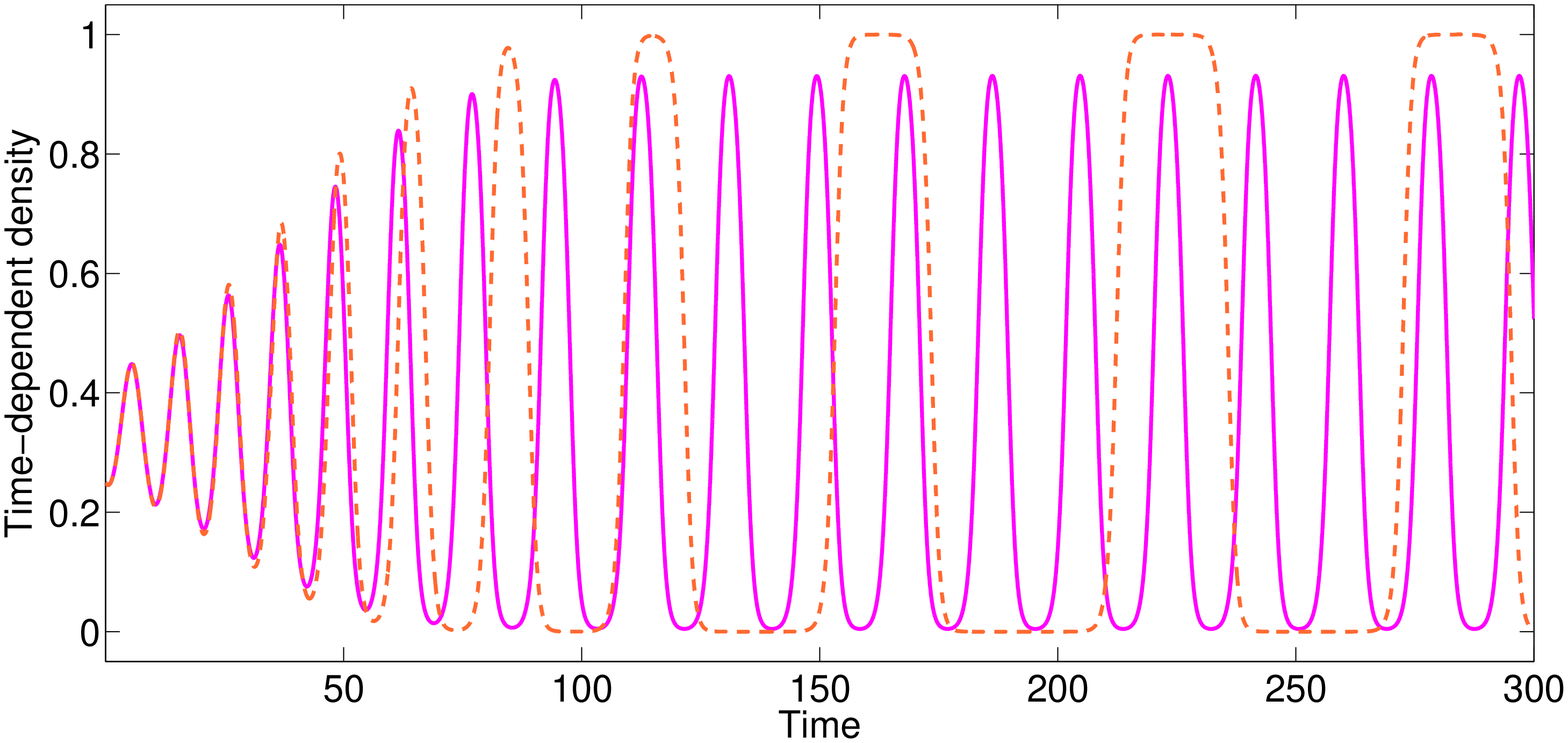}
\includegraphics[ height=4.0cm,clip=]{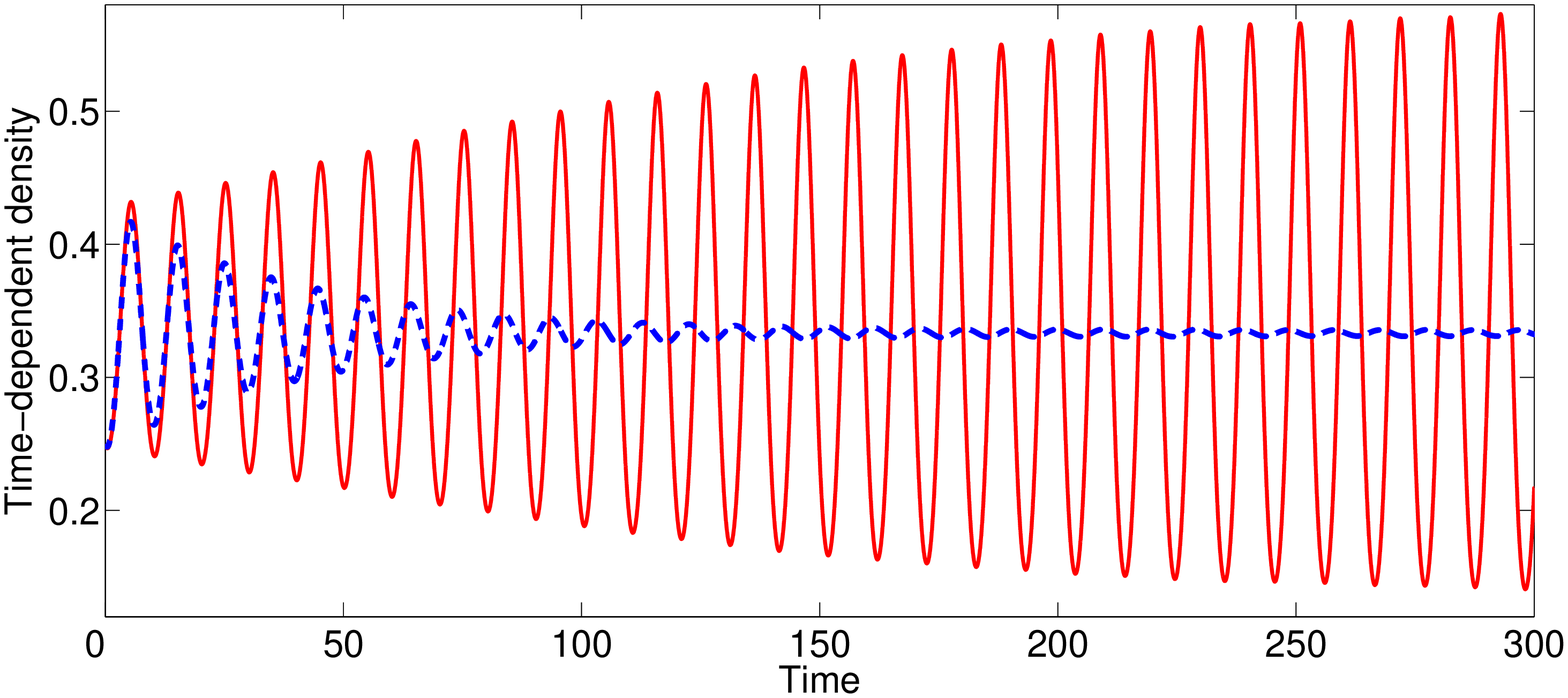}
\caption{ {\it (Color online)}. Typical plots of the time-dependent density (relative abundance) $a(t)$ of species $A$, obtained from Eqs.~(\ref{RE1})
for $\epsilon=1.225$ and different values of $\mu$ (with $a(0)=0.25, b(0)=0.40$). Here, $\mu_c=0.0125$ (see text). Left: In the absence of mutation rate ($\mu=0, \lambda=0.0375$), the fixed point ${\bm s}^*$ is unstable and the density (brown, dashed) jumps from  $0$ to  $1$. For very low values of the mutation rate, here  $\mu=0.001$ ($\lambda=0.0345$),  ${\bm s}^*$ is also unstable and $a(t)$  oscillates regularly (magenta, solid) with an amplitude that approaches the extreme values $0$ and $1$, but with
 a much shorter period  than in the case $\mu=0$. 
 Right: For $\mu=0.01$ ($\lambda=0.0075$), the fixed point ${\bm s}^*$ is still unstable and $a(t)$ oscillates regularly about $a^{*}=1/3$ (red, solid),  with a period $T\approx 10$.
When $\mu=0.02>\mu_c$ ($\lambda=-0.0225$), the densities 
exhibit exponentially damped oscillations and converge towards the fixed point value  $1/3$ (blue, dashed).
\label{oscillations}}
\end{center}
\end{figure}
\section{The Rate Equations}
Here, we consider the dynamics of the model RPSM  in the mean-field limit, where $N\to \infty$. When the population is well-mixed (homogeneous) and every pair of random individuals has the same probability to interact, demographic noise can be neglected and the system's dynamics is aptly described by the rate equations
\begin{eqnarray}
\label{RE1}
\dot{a}&=& a\left[c-\epsilon b -(1-\epsilon)\left\{ab+bc+ac\right\}\right] + \mu(1-3a)\nonumber\\
\dot{b}&=& b\left[a-\epsilon c -(1-\epsilon)\left\{ab+bc+ac\right\}\right] + \mu(1-3b)\\
\dot{c}&=& c\left[b-\epsilon a -(1-\epsilon)\left\{ab+bc+ac\right\}\right] + \mu(1-3c)\nonumber
\end{eqnarray}
As the system is comprised of three species, the sum of the population density is conserved, i.e. $a(t)+ b(t) + c(t)=1$. Clearly, this constant of motion allows, say,  to set $c(t)=1-a(t)-b(t)$  and reduces  (\ref{RE1}) to a  system  with only two variables (here, $a(t)$ and $b(t)$). Thereafter, the properties of these equations are studied and it will be shown  that  mutations yield
a new region in the parameter space characterized by a stable limit cycle.

It is known that in the absence of mutations ($\mu=0$) the REs admit one interior fixed point, ${\bm s^*}=(a^*,b^*,c^*)=(1/3,1/3,1/3)$, and three absorbing states ($(1,0,0)$, $(0,1,0)$ and $(0,0,1)$). The resulting cyclic dynamics gives rise to three kinds of behavior~(\cite{Hofbauer,Nowak,szabo-2007-446}): (i)  When $\epsilon<1$,  ${\bm s^*}$ is the system's only  attractor, it is  globally stable  and the trajectories in the phase portrait spiral towards it (${\bm s^*}$ is a focus).
 (ii) When  $\epsilon>1$, the interior rest point ${\bm s^*}$ becomes unstable and the flows in the phase portrait 
form a heteroclinic cycle connecting each of the absorbing fixed points (saddles) at the boundary of the phase portrait.
Clearly, as any fluctuations cause the rapid extinction of two species, the resulting oscillations are non-robust~(\cite{may-1975-29,durrett-1994-46}).
 (iii)  When $\epsilon= 1$,  $\bar \pi$ vanishes (zero-sum game)  and the interior rest point ${\bm s^*}$ is  marginally stable (center). In this case, the quantity $a(t)b(t)c(t)$ is a constant of motion and the trajectories in the phase portrait  form  closed orbits around ${\bm s^*}$, set by the initial conditions and not robust against noise  (see e.g.~(\cite{reichenbach-2006-74})).

In the presence of mutations, solving Eqs.~(\ref{RE1}) with $\dot{a}=\dot{b}=\dot{c}=0$, 
one finds that ${\bm s^*}=(1/3,1/3,1/3)$ is the
only (interior) fixed point of the system. The absence of absorbing points when $\mu>0$ indicates that the model under consideration does not yield heteroclinic cycles.
In fact, as shown below, in the presence of small mutation rate the heteroclinic cycles are replaced by stable cycles resulting from a Hopf bifurcation (region (b) in Fig.~\ref{bifdiagram})
and yield persistent oscillatory dynamics of all species, which is a feature
 of biological relevance.

\subsection{Linear stability analysis}
\begin{figure}    
\begin{center} 
\includegraphics[scale=.68]{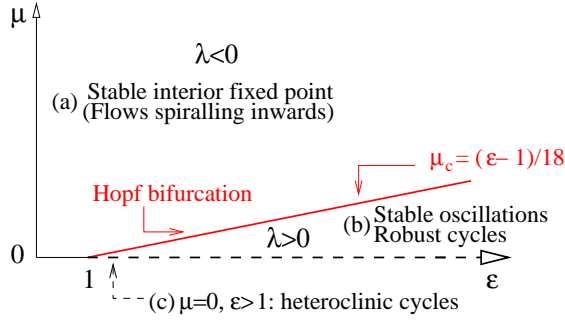}
\caption{{\it (Color online).}  Bifurcation diagram in the parameter space of the (deterministic) model RPSM. 
The regions (a) and (b) are separated by the critical mutation rate $\mu_c=(\epsilon-1)/18>0$.
In region (a),  $\mu>
\mu_c$ for $\epsilon>1$ and $\mu\geq 0$ for $\epsilon \leq 1$,
 ${\bm s}^{*}$ is a stable focus. Below the line $\mu_c(\epsilon)$,
 in region (b) where $0<\mu<\mu_c$, the system undergoes a Hopf bifurcation (at $\mu=\mu_c$) and the dynamics is characterized by a limit cycle  and stable oscillations.
In region (c), where $\mu=0$ (no mutations) and $\epsilon>1$ (dashed line), the REs (\ref{RE1}) yield heteroclinic cycles along with periodic oscillations that are unstable against demographic fluctuations.  
\label{bifdiagram}}  
\end{center}
\end{figure}
To investigate the properties of REs~(\ref{RE1}), it is useful to introduce the 
variables ${\bm x}=(x_A, x_B)$:
\begin{eqnarray}
\label{x}
x_A=a-a^*=a-\frac{1}{3} \quad \text{and} \quad x_B=b-b^*=b-\frac{1}{3},
\end{eqnarray}
 which measure the deviations from the interior fixed point ${\bm s^*}$. We notice that $x_C=c-c^*=-(x_A+x_B)$.  As a first step to gain some insight into the dynamics described by (\ref{RE1}),  it is useful to perform a linear stability analysis. To linear order in terms of ${\bm x}=(x_A, x_B)$), the REs (\ref{RE1}) can be rewritten as $\dot{\bm x}= \mathcal{A}({\bm s^*}){\bm x}$, 
where $\mathcal{A}({\bm s^*})$ is 
the Jacobian matrix at ${\bm s^*}$ whose (complex conjugate) eigenvalues which are $ \lambda_{\pm}= \lambda \pm i\,\omega_0$, with
\begin{equation} 
\label{L}
  \lambda=\frac{\epsilon -1}{6}-3\mu \quad \text{and} \quad \omega_0=\frac{1+\epsilon}{2\sqrt{3}}.
\end{equation}
General results therefore guarantee that  ${\bm s}^*$ is (globally) stable  when $\lambda<0$, i.e. for $18\mu>\epsilon-1$, whereas 
${\bm s}^*$ is unstable when $\lambda>0$ and a {\it Hopf bifurcation} occurs at the critical value $\mu=\mu_c(\epsilon)$ and $\epsilon>1$ is kept fixed, with
\begin{equation}
 \label{muc}
\mu_c\equiv
     \frac{\epsilon -1}{18}>0,
\end{equation}
 The bifurcation diagram of the model is thus summarized in Fig.~\ref{bifdiagram}. Such a diagram clearly illustrates the effect of mutations and is characterized by three regions.  In region (b), where 
$0<\mu<\mu_c$, ${\bm s}^*$ is unstable and there is  a Hopf bifurcation at $\mu=\mu_c$ (as demonstrated below) which leads to the coexistence of all species with stable oscillations of their densities (right panel of Fig.~\ref{oscillations}, red/solid).  On the other hand, in region (a), where $\mu> \begin{cases} 
\mu_c & \; \text{, for} \quad \epsilon>1\\
0 & \;              \text{, for} \quad  \epsilon 
\leq 1        \end{cases},$  i.e. when $\lambda<0$, the densities are exponentially
damped, with an amplitude ${\rm A}_{\rm amp}(t)\approx e^{-|\lambda|t}$  that vanishes near $\bm{s}^*$  (right panel of Fig.~\ref{oscillations}, blue/dashed).
In the absence of mutations and when $\epsilon>1$ (region (c) in Fig.~\ref{bifdiagram}), one recovers the scenario characterized by heteroclinic cycles (left panel of Fig.~\ref{oscillations}, brown/dashed). In this case, the density of each species  oscillates, with a large period, between the extreme values $0$ (absence of a given species) and $1$ (presence of only one  species). These
 oscillations are not
 robust against demographic noise: chance fluctuations  will unavoidably and quickly drive the system into an absorbing state. 
When $\lambda=0$, the fixed point  ${\bm s}^*$ is a center and its stability is determined by nonlinear terms.
Below, we shall consider the dynamics to third order about  ${\bm s}^*$ and show that it is stable when $\lambda=0$ and $\mu>0$. Only in the marginal case $\lambda=\mu=0$, 
${\bm s}^*$ is marginally stable   (see e.g. (\cite{reichenbach-2006-74})).
\subsection{Hopf bifurcation and limit cycle}
As qualitatively discussed above, at mean-field level, the most interesting effect of mutations  is the emergence of robust oscillatory behavior (stable limit cycle) resulting from a Hopf bifurcation occurring at low mutation rates (right panel of Fig.~\ref{oscillations}, red/solid). To study the properties of the resulting limit cycle and take advantage of the system's symmetry around  ${\bm s}^*$, it is convenient to perform the linear transformation ${\bm x} \to {\mathcal  S}{\bm x}\equiv {\bm y}$, where ${\mathcal  S}=
\begin{pmatrix}
1 & \frac{1}{2} \\
0 & \frac{\sqrt{3}}{2}
\end{pmatrix}$.
As well known from bifurcation theory~(\cite{Wiggins,Floquet}), one then Taylor expands 
the REs (expressed in the variables ${\bm y}={\mathcal  S}{\bm x}$)
to third-order around ${\bm s}^*$   and recasts the resulting rate equations into the Hopf bifurcation normal form. Hence, in polar coordinates $(r,\theta)$,  the  normal form of the REs (truncated to third order) reads
\begin{eqnarray}
\label{polar-NF}
\dot{r}&=& r(\lambda + \beta r^{2}) \nonumber\\
\dot{\theta}&=& \omega_0 - \alpha r^{2},
\end{eqnarray}
where
\begin{eqnarray}
\label{gamma}
\alpha &=&\frac{18\omega_0(1+2\sqrt{3}\omega_0)}{7(1+\epsilon^2)+\epsilon(13-9\mu)+9\mu(1+9\mu)}\\
\beta&=& 1-\epsilon - \left(\frac{6\lambda(1+2\epsilon \sqrt{3} \omega_0)}{7(1+\epsilon^2)+\epsilon(13-9\mu)+9\mu(1+9\mu)}\right).\nonumber
\end{eqnarray}
One readily recognizes that  (\ref{polar-NF}) yield a {\it supercritical Hopf bifurcation} when $\lambda>0$ (for $0<\mu<\mu_c(\epsilon)$).
In fact, the parameter $\beta$ is the first Lyapunov coefficient and is negative when $\lambda>0$ and the mutation rate $\mu$ is small (i.e. $\mu/\epsilon \ll 1$)~(\cite{Wiggins,Floquet,Strogatz}). 

\begin{figure}    
\begin{center} 
\includegraphics[scale=0.5]{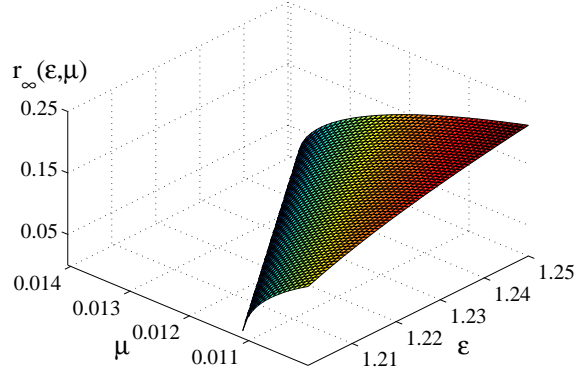}
\caption{{\it (Color online).} Plot of $r_{\infty}=\sqrt{\lambda/|\beta|}>0$ for $\mu>\mu_c$  as function of $\epsilon$ and $\mu$ according to (\ref{L}) and (\ref{gamma}). We notice that $r_{\infty}$ increases when $\epsilon$ is raised and $\mu$ is lowered.  \label{rinf}}
\end{center}
\end{figure}

It is insightful to solve the radial component of the Eq.~(\ref{polar-NF}), whose solution is $r(t) = \frac{r(0) \;e^{\lambda t}}{\sqrt{1-r(0)\frac{\beta}{\lambda}(e^{2\lambda t}-1)}}$. Thus, when $\lambda>0$, the interior fixed point ${\bm s}^*$ is unstable and the long-time behavior is $r(t) \simeq r_{\infty} \left(1-\frac{e^{-2\lambda t}}{2}\left\{\frac{\lambda -r^2(0)|\beta|}{r^2(0)|\beta|}\right\}\right)$,
where
\begin{eqnarray}
\label{R}
r_{\infty}= \sqrt{\frac{\lambda}{|\beta|}}.
\end{eqnarray}
The dependence of $r_{\infty}$ on the parameters $\epsilon$ and $\mu$ is illustrated in Figure~\ref{rinf}, where it is shown that $r_{\infty}$ increases when $\epsilon$ is raised and  $\mu$ lowered.
Thus, it follows from Eqs. (\ref{polar-NF},\ref{R})  that the REs (\ref{RE1}) indeed yield a {\it stable limit cycle} $\bar{\bm \sigma}(t)=(\bar{a}(t), \bar{b}(t))$ (with $\bar{c}(t)=1-\bar{a}(t)-\bar{b}(t)$) of radius $r_{\infty}$, period $T$ and frequency $\omega$, where
\begin{eqnarray}
\label{period}
T = \frac{2\pi}{\omega}, \quad \text{with} \quad \omega= \omega_0 -  \frac{\lambda \alpha}{|\beta|} \
\end{eqnarray}
We notice that the linear terms contribute to the frequency $\omega$ through $\omega_0$ (natural frequency), while nonlinearity gives rise to an additional  contribution ($-\lambda \alpha/|\beta|$). 
According to (\ref{polar-NF}), the limit cycle $\bar{\bm \sigma}(t)$ is approached exponentially fast in time, $r(t)-r_{\infty}\propto e^{-2\lambda t}$.

\begin{figure}    
\begin{center} 
\includegraphics[scale=0.44]{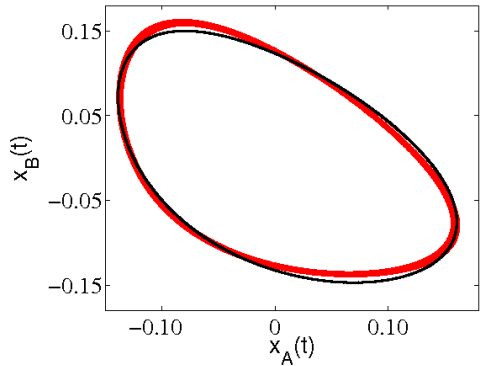}
\caption{{\it (Color online).} With $\bar{x}_A(t)=\bar{a}(t) -\frac{1}{3}$ and $\bar{x}_B(t)=\bar{b}(t) -\frac{1}{3}$, comparison of the limit cycle  obtained from the numerical solution of (\ref{RE1})  (red/thick, here $550 \leq t \leq 1000$)   
with the predictions of  (\ref{polar-NF}) [black/thin].  The parameters  and initial conditions   are $\epsilon=1.2,~\mu=0.01, x_A(0)=0.025, x_B(0)=0.040$, here $\lambda=1/300$.\label{comp}}
\end{center}
\end{figure}

As they result from a third-order expansion, the predictions of (\ref{polar-NF}) are  accurate for mutation rates close to the critical value  $\mu_c$ (with $\epsilon$ fixed), i.e. for ``small'' positive values of $\lambda$ and $r_{\infty}$ as illustrated in Fig.~\ref{comp}. 

To conclude this section, we also consider the case where $\lambda<0$ and notice that   Eqs.~(\ref{polar-NF}) thus yield
 $r(t)\stackrel{t \to \infty}{\longrightarrow} e^{-|\lambda|t}$. This implies that
${\bm s}^*$ is approached in an oscillatory manner with an exponentially damped amplitude, namely
$x_A(t)\propto  e^{-|\lambda|t}\; \left(
\cos{(\omega_0 t )} -\sin{(\omega_0 t)}/\sqrt{3}\right)$ and $x_B(t) \propto  e^{-|\lambda|t}\; \sin{(\omega_0 t)}$.
 Furthermore, in the marginal case $\lambda=0$ (with $\mu>0$),  ${\bm s}^*$ is still stable but approached in a slow oscillatory manner. In fact, it follows from (\ref{polar-NF}) that in  this case $r(t)\sim t^{-1/2}$, which implies that 
 the
oscillations are damped
algebraically as $t^{-1/2}$. Indeed, for $\lambda=0$ one finds $x_A(t)\stackrel{t \to \infty}{\longrightarrow} \left[\cos{(\omega_0 t)} -\frac{1}{\sqrt{3}}
\sin{(\omega_0 t)}\right]/(6\sqrt{\mu t})$ and $x_B(t) \stackrel{t \to \infty}{\longrightarrow} \sin{(\omega_0 t)}/(3\sqrt{3\mu t})$.

\section{The Stochastic Nonlinear Dynamics}
\label{stochastic}
\begin{figure}  
\begin{center} 
\includegraphics[scale=0.35]{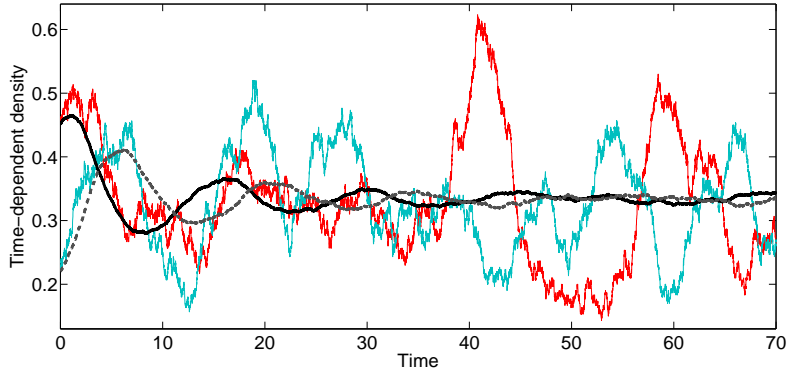}
\caption{{\it (Color online).} Erratic oscillations and
stochastic dynamics of the model RPSM in the region (a) of the parameter space, with  $\epsilon=0.5, \mu=0.01$ and $N=300$ (initially $N_A=135$ and  $N_B=66$). In single realizations, the  time-dependent densities (relative abundances) of species $A$ (thin, red/dark grey) and $B$ (thin, cyan/light grey)  are characterized by persistent stochastic oscillations resulting from a resonance amplification occurring at frequency $\Omega^*\simeq \omega_0 \simeq 0.43$ (see text). When they are sample-averaged (here over $200$ replicates), the mean densities (solid and dashed thick curves for species $A$ and $B$, respectively) display  damped oscillations approaching the value $1/3$, in agreement with (\ref{RE1}).\label{StochRes}}
\end{center}
\end{figure}
We so far  have  focused on the deterministic  description of the model in the mean-field limit. However,
 as virtually all real systems are of finite size, $N<\infty$, and made of discrete entities, they are influenced by stochastic fluctuations.
Here, adopting an  individual-based approach, we discuss how demographic noise alters the deterministic properties of the model RPSM.

\begin{figure} 
\begin{center} 
\includegraphics[scale=0.35]{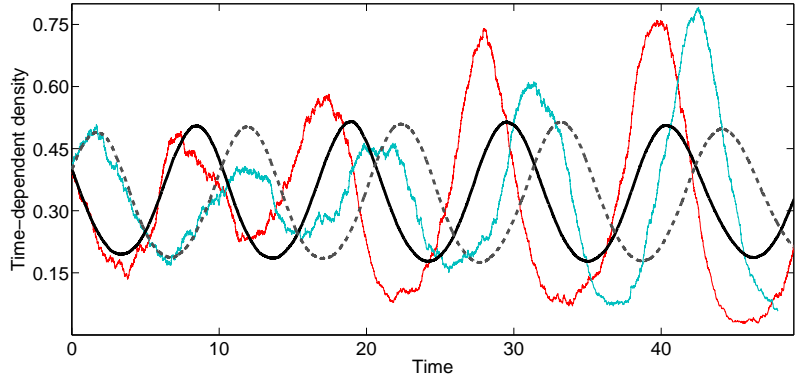}
\caption{{\it (Color online).} Stochastic dynamics of the model RPSM in the region (b) of the parameter space with $\epsilon=1.25, \mu=0.008$ and $N=600$ (initially $N_A=N_B=240$). We notice that the averaged (over $250$ replicates) time-dependent densities of species $A$ (thick, solid) and $B$ (thick, dashed) are characterized by oscillations of non-vanishing amplitude. The erratic trajectories in red/dark grey and cyan/light grey are single realizations (no sample averaging) of the density of species $A$ and $B$, respectively. \label{StochRes_regb}}
\end{center}
\end{figure}

\begin{figure} 
\begin{center} 
\includegraphics[scale=0.5]{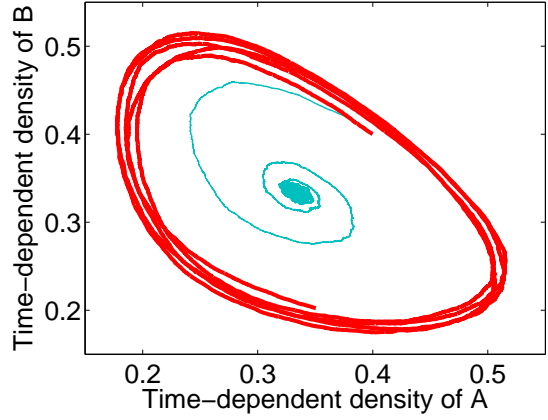}
\caption{{\it (Color online).} Phase portrait of the model RPSM 
in the presence of stochastic fluctuations for $\epsilon=1.25, N=600$, with  $\mu= 0.008$ (thick, red/dark grey) and $\mu= 0.05$ (thin, cyan/light grey), initially $N_A=N_B=240$. The  trajectories have been obtained  from a sample-average over $250$ replicates. When   $\mu= 0.008$, the interior fixed point ${\bm s}^*$ is unstable and each trajectory in the phase portrait forms a ``phase-remembering quasi-cycle''of erratic 
non-vanishing  amplitude and phase orbiting around  $\bar{\bm \sigma}$. When $\mu= 0.05$, the flow spirals towards the stable interior fixed point ${\bm s}^*$ and erratically wanders in its vicinity forming a ``phase-forgetting quasi-cycle'' (see text).
\label{PhasePortStoch}}
\end{center}
\end{figure}

Before mathematically studying the model RPSM in the presence of demographic noise, it is worth gaining some 
insight into its stochastic dynamics. As explained below, the latter has been 
implemented by a birth-death process
simulated according to the Gillespie algorithm~(\cite{gillespie1,gillespie2}). Some characteristic results are reported in Figs.~\ref{StochRes}-\ref{PhasePortStoch}. The behavior corresponding to the regime (a) of the parameter space (see Fig.~\ref{bifdiagram}) is illustrated in Fig.~\ref{StochRes}, where demographic noise  causes erratic oscillations forming ``(phase-forgetting) quasi-cycles''~(\cite{Nisbet}) in the phase portrait (see also Fig.~\ref{PhasePortStoch}, cyan/light grey trajectory). When the population size $N$ is large and sample-averaged over many replicates, the amplitude of the stochastic oscillations decreases and the predictions of the rate equations are approached (thick curves in Fig.~\ref{StochRes}), but fully recovered only in the mean-field limit  $N\to \infty$.
 The stochastic dynamics in the region (b) of the parameter space, where  ${\bm s}^*$ is unstable, is illustrated in Fig.~\ref{StochRes_regb} and is characterized by persistent oscillations with  erratic, but non-vanishing, amplitude and phase. As a result, instead of a perfectly closed orbit the trajectories in the phase portrait form a perturbed limit cycle, also called ``phase-remembering quasi-cycle''~(\cite{Nisbet}), as shown in  Fig.~\ref{PhasePortStoch} (red/dark grey trajectory). Figs.~\ref{StochRes} and \ref{StochRes_regb} show that demographic noise 
affects considerably each single realization in the region (a) of the parameter space, where the predictions of the REs are replaced by oscillations of non-vanishing amplitude, while it  perturbs the cyclic behavior in region (b).
Below, we show that the ``quasi-cycles'' arising in region (a) of Fig.~\ref{bifdiagram} at low mutation rate  
 result from a resonance amplification  mechanism~(\cite{McKane1}). We will also see 
how noise affects the system's attractors that can be regarded as
minima of an effective potential well from which it takes an enormous amount of time to escape~(\cite{Gardiner,Kubo,Dykman,Volovik}).

To describe the stochastic nonlinear dynamics of the system, we adopt an individual-based approach and consider an urn model~(\cite{Feller}) comprising a total of $N=N_A+N_B+N_C$ individuals, $N_A$ are of species $A$, $N_B$  of species $B$, and $N_C$ of species $C$. Up to fluctuation of order ${\cal O}(N^{-1/2})$~\footnote{In fact, the number of individuals of species $A$ is related to $a$ by $N_{A}=aN+{\cal O}(\sqrt{N})$ and similarly with the other species.} (see below), the densities of species $A$, $B$ and $C$ are respectively $a=N_A/N$, $b=N_B/N$
and $c=N_C/N=1-a-b$. The stochastic dynamics is here implemented according to the frequency-dependent Moran process (fMP)~(\cite{Nowak-bis,Nowak}) (see also (\cite{moran-1958-54,Blythe})). The latter is a Markovian process in which, at each time step, two individuals -- say one of species $A$ and another of species $B$ -- are randomly selected and then reproduce according to their fitness, here  proportional to the individual's average payoff.
Once the density-dependent average payoffs  are determined, one of the individual, say $A$, reproduces with a rate $T^{B\to A}$ and the offspring replaces a randomly chosen individual of the other species ($B$ in this example). The fMP 
therefore conserves the total size $N$ of the population and 
can be regarded as a random-walk with hopping rates depending on each species average payoff (fitness).
 While there are various ways of implementing the dynamics of the fMP, we here consider that in the absence of mutations and large (yet finite) population size
 the transition $T^{B\to A}$ from $B$ to $A$, say, is given by $[1+\left(\pi_A -\bar{\pi}  \right)]\, ab$. This expression comprises a contribution proportional to the average payoff differences $(\pi_A -\bar{\pi})$, that accounts for selection, supplemented by a constant (set to $1$)
accounting for the background random noise~(\cite{Nowak-bis,Nowak}). The multiplying factor $ab$ encodes the probability (when $N\gg 1$) that two individuals of species $A$ and $B$ interact. 
 The effect of mutations is thus taken into account by adding a linear term, which finally leads to the following transition rate: $T^{B\to A}({\bm s}) = \left(1+\{\pi_A -\bar{\pi}\} \right)\, ab + \mu b$. More generally, we consider the following transition rates:
\begin{eqnarray}
\label{TR}
T^{i\to j}({\bm s}) &=& \left(1+\{\pi_j -\bar{\pi}\} \right)\, s_is_j + \mu s_i, 
\end{eqnarray}
with $i\neq j \in (A,B,C)$ and $\pi_A=c-\epsilon b$, $\pi_B=a-\epsilon c$, $\pi_C=b-\epsilon a$ and $\bar{\pi}=(1-\epsilon)(ab+bc+ac)$.

The stochastic description of the system is encoded in the probability $P(N_A,N_B;t)$ of having
$N_A$ and $N_B$ individuals of species $A$ and $B$, respectively, at time $t$. The quantity $P(N_A,N_B;t)$  obeys the master equation  associated with the above fMP (see, e.g., (\cite{Gardiner,Ewens}))
and  specified by the transition rates (\ref{TR}). The Markovian process defined by the transition rates (\ref{TR}) has been simulated
using the Gillespie algorithm~(\cite{gillespie1,gillespie2}) which was initially introduced to simulate chemical systems through their ``microscopic reactions''. In the vicinity of ${\bm s}^{*}$ and
for large (yet finite) population size, the master equation of the above fMP can 
be aptly described within a generalized diffusion approximation for the probability density $P({\bm x},t)$, where ${\bm x}=(x_A,x_B)$ and $x_j\equiv \frac{N_j}{N}-\frac{1}{3},\; j \in (A,B)$ 
are considered continuous variables (about which we expect fluctuations of order $N^{-1/2}$, see below). The temporal development of  $P(x_A,x_B;t)$ is thus described by a forward Kolmogorov equation (FKE), or Fokker-Planck equation, resulting from a size-expansion of the master equation (see, e.g., (\cite{Gardiner,Ewens,Nisbet,traulsen-2005-95})). Performing a (linear) van Kampen expansion (\cite{VanKampen}) 
about the fixed point ${\bm s}^{*}$
in the continuum limit (assuming that $N$ is large but finite), the standard procedure  
(see, e.g., (\cite{Gardiner,Risken,reichenbach-2006-74})) leads to the following FKE:
\begin{eqnarray}
\label{FPE}
\partial_t P({\bm x},t)=-\partial_{x_i}\left[ x_j {\cal A}_{ij}({\bm s}^{*})P({\bm x},t) \right] + \frac{1}{2} {\cal B}_{ij}({\bm s}^{*}) \partial_{x_i} \partial_{x_j}  P({\bm x},t),
\end{eqnarray}
where we have adopted the summation convention on repeated indices $i,j \in (A,B)$. In Eq.~(\ref{FPE}) the drift terms $({\cal A}({\bm s}^{*})\,{\bm x})_i$ are obtained from the Jacobian matrix ${\cal A}$ of the REs (\ref{RE1}) at ${\bm s}^{*}$ (see Sec. 3.1), while the diffusion matrix ${\cal B}$ is defined by (see, e.g.,~(\cite{Claussen}))
\begin{eqnarray}
\label{B}
{\cal B}_{AA}({\bm s}^{*})&=&T^{B\to A}({\bm s}^*)+T^{A\to B}({\bm s}^*)+T^{C\to A}({\bm s}^*)+T^{A\to C}({\bm s}^*)=\frac{4(1+3\mu)}{9N}
\nonumber\\{\cal B}_{BB}({\bm s}^{*})&=&
T^{A\to B}({\bm s}^*)+T^{B\to A}({\bm s}^*)+T^{C\to B}({\bm s}^*)+T^{B\to C}({\bm s}^*)
=\frac{4(1+3\mu)}{9N}\\
{\cal B}_{AB}({\bm s}^{*})&=&{\cal B}_{BA}({\bm s}^{*})=-\left\{T^{A\to B}({\bm s}^*)+T^{B\to A}({\bm s}^*)\right\}
=-\frac{2(1+3\mu)}{9N}. \nonumber
\end{eqnarray}
As well known, the FKE (\ref{FPE}) is equivalent to the following set of linear stochastic (Langevin) differential equations with Gaussian white noise vector ${\bm \eta}=(\eta_A, \eta_B)$~(\cite{Gardiner,Risken}):
\begin{eqnarray}
\label{SDE}
\dot{x}_A &=& {\cal A}_{AA}({\bm s}^{*}) \; x_A + {\cal A}_{AB}({\bm s}^{*}) \; x_B +\eta_A\nonumber\\
\dot{x}_B &=& {\cal A}_{BA}({\bm s}^{*}) \; x_A +  {\cal A}_{BB}({\bm s}^{*}) \; x_B +\eta_B, 
\end{eqnarray}
 with $\langle \eta_i \rangle=0$ and covariance matrix $\langle \eta_i (t) \eta_j (t')\rangle = {\cal B}_{ij}({\bm s}^{*}) \, \delta(t-t')$, where $i,j \in (A,B)$ and $\langle ... \rangle$ denotes the ensemble average over a large number of replicates.
As anticipated, it follows from (\ref{B},\ref{SDE})  that the noise strength is  $\propto N^{-1/2}$ and yields fluctuations of 
 order ${\cal O}(N^{-1/2})$ around the deterministic values of $x_A$ and $x_B$.

 Below, the discussion of  stochastic effects on the evolutionary dynamics of the model are centred on two aspects: (i) we show how demographic noise leads to the emergence of quasi-cycles; (ii) and we compute the average time for a system with  the same initial density of each species to ``escape'' from the interior rest point and reach a given cycle.

\subsection{Quasi-cycles and noise-induced resonance amplification}
\begin{figure} 
\begin{center} 
\includegraphics[scale=0.29]{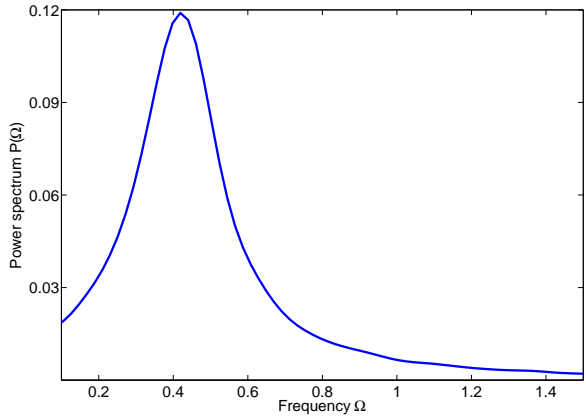}
\caption{{\it (Color online).} Power spectrum $P(\Omega)=\langle |{\widetilde {\bm x}}(\Omega)|^{2} \rangle$ of the model RPSM with $\epsilon=0.5, \mu=0.01$ and $N=300$ obtained from numerical simulations and by sample averaging over $200$ realizations. We notice  an isolated peak at characteristic frequency $\Omega^*\simeq 0.43$, in full agreement with the analytical prediction (\ref{omegastar}).\label{PW}}
\end{center}
\end{figure}
In this section, we are especially interested in the stochastic dynamics in region (a) of the phase parameter, where $\lambda<0$ and  ${\bm s}^*$ is a stable fixed point.
Here, we investigate 
one of the most intriguing effects of intrinsic noise, 
which  yields persistent erratic oscillations around ${\bm s}^*$ (see Fig.~\ref{StochRes}) and therefore considerably alters the predictions of the REs (\ref{RE1}).
In fact, while one could naively expect  only small corrections (of order $N^{-1/2}$, see Eqs.~(\ref{B},\ref{SDE}))  to the deterministic predictions, 
 it has been suggested that demographic noise can be sufficient to perturb the stationary state predicted by the mean-field  analysis and to produce persistent erratic oscillations~(see e.g.~(\cite{Bartlett1,Bartlett2})). This is indeed illustrated by the numerical simulations reported in Fig.~\ref{StochRes}. 
It has recently been shown that such a behavior, often referred to as {\it quasi-cyclic}~(\cite{Nisbet}), results from a resonant amplification of demographic (intrinsic) fluctuations~(\cite{McKane1}). To analytically demonstrate this phenomenon, we shall compute the  {\it power-spectrum} $P(\Omega)$ of the system. The power-spectrum is one of the most useful tools to look for oscillations in noisy data and is related to the auto-correlation functions of (stationary) systems~(\cite{Nisbet,Gardiner}).
Introducing ${\widetilde {\bm x}}(\Omega)\equiv \frac{1}{2\pi}\int_{-\infty}^{\infty} dt\; e^{-i\Omega t} \, {\bm x}(t)$, the Fourier transform of ${\bm x}(t)$, the power spectrum is given by the following ensemble average: $P(\Omega)=\langle |{\widetilde {\bm x}}(\Omega)|^{2} \rangle=2\langle |{\widetilde x_A}(\Omega)|^{2} \rangle=2\langle |{\widetilde x_B}(\Omega)|^{2} \rangle$. Taking the Fourier transform of Eqs.~(\ref{SDE}) and then averaging the square modulus of ${\widetilde {\bm x}}(\Omega)$ one finds:
\begin{eqnarray}
\label{Pw}
P(\Omega)&=&
\frac{8(1+3\mu)}{9N}\quad \frac{\Omega_0^2 + \Omega^2}{\left(\Omega^2 - \Omega_0^2 \right)^2 + \left(2\lambda \Omega\right)^2}, \nonumber\\  \text{with} \quad
9\Omega_0^2 &\equiv& 1+ 2\sqrt{3}\epsilon\omega_0 +9\mu(9\mu+1-\epsilon).
\end{eqnarray}
 For low mutation rate $\mu$ (i.e. when $\mu/\epsilon \ll 1$), $\Omega_0^2>2\lambda^2$ and the power spectrum $P(\Omega)$ has a single peak (for $\Omega>0$) at the characteristic frequency
\begin{eqnarray}
\label{omegastar}
\Omega^*&=& \Omega_0 \left( 2\sqrt{1-\left(\frac{\lambda}{\Omega_0}\right)^2}
-1
\right)^{1/2},
\end{eqnarray}
as illustrated in Fig.~\ref{PW}.
When the mutation rate $\mu$ is low, the value of $\Omega^*$ is very close to  (yet different from) $\omega_0$ and  $\Omega_0$ (\ref{L}, \ref{Pw})~\footnote{E.g., for $\epsilon=0.5$ and $\mu=0.01$, one finds: $\omega_0 \simeq 0.4330$, whereas $\Omega^* \simeq 0.4328$ and $\Omega_0 \simeq 0.4476$.}.
When $(\lambda/\Omega_0)^2\ll 1$, the above expression simplifies to give $\Omega^*= \Omega_0(1+O(\lambda^2/\Omega_0^2))$. 
It is also worth noticing that at the onset of the Hopf bifurcation, i.e. when $\lambda=0$ (with $\mu>0$), the denominator of $P(\Omega)$ vanishes for $\Omega=\Omega_0$ and in this case the peak in the power spectrum is replaced by a pole at $\Omega=\Omega^*=\Omega_0$.

While on general grounds (law of large numbers~(\cite{Feller})) one expects oscillations of the order $N^{-1/2}$, 
the presence of a peak in the power spectrum corresponds to a noise-induced amplification due to a resonance effect. In fact, there is a resonant behavior when there exists a particular frequency  for which the denominator of the power spectrum $P(\Omega)$ is small. Here, the denominator of the power spectrum  takes its minimal value 
at 
\begin{eqnarray}
\label{omega-res}
\Omega_{\rm res}= \sqrt{\Omega_0^2 -2\lambda^2},
\end{eqnarray}
which simplifies when $(\lambda/\Omega_0)^2\ll 1$ and also gives  $\Omega_{\rm res}=\Omega_0(1+O(\lambda^2/\Omega_0^2))$. In such a regime, where $\Omega^*\simeq \Omega_{\rm res} \simeq \Omega_{0}$, there is 
a very suggestive explanation of the noise-induced resonant amplification by comparison with a simple mechanical system~(\cite{McKane1}). Indeed, one can consider a linear damped harmonic oscillator of natural frequency $\Omega_0$, with damping constant $\zeta$,  driven by a force oscillating with frequency $\Omega$. One can thus show that such a mechanical system oscillates at frequency $\Omega$, with  amplitude ${\rm A_{amp}}(\Omega)\propto \left[(\Omega^2-\Omega_0^2) + (\zeta\Omega)^2\right]^{-1}$ and yields a resonant frequency $\Omega_{\rm res}=\sqrt{\Omega_0^2 -\frac{\zeta^2}{2}}$. Thus, the quantity $2\lambda$ can be interpreted as the damping constant $\zeta$ of the mechanical system. Therefore, provided that $2\lambda$ is much smaller than $\Omega_0$ (i.e. $\lambda^2/\Omega_0^2 \ll 1$), the model's dynamics is essentially the same  as for a linear damped harmonic oscillator with a resonant frequency  $\Omega_{\rm res}=\Omega_0$.
Yet, while the driving frequency has to be tuned to achieve resonance in the mechanical system, this is not the case for the noisy system that we are considering. In fact, as the model RPSM is driven by internal (Gaussian) white noise (demographic stochasticity) which covers all frequencies, the resonant frequency of the system will be excited without need of any external tuning.

Other fruitful quantities to measure repetitiveness and (quasi-)periodicity in a fluctuating population are the autocorrelation functions $\langle x_{A}(\tau+t) x_{A}(\tau)\rangle = \langle x_{B}(t+\tau) x_{B}(t)\rangle$ and  $\langle x_{A}(\tau + t) x_{B}(\tau)\rangle = -\langle x_{B}(\tau + t) x_{A}(\tau)\rangle$. The latter are related to the the power spectrum by the Wiener-Khinchin theorem and can be computed by the Fourier transform of $P(\Omega)$~(\cite{Gardiner}), which yields (for $t\geq \tau \gg 1$)
\begin{eqnarray}
\label{autocorr}
\langle x_{A}(\tau+t) x_{A}(\tau)\rangle&=& \frac{4(1+3\mu)}{3N}\; \frac{e^{-|\lambda|t}}{|\lambda|}\, \cos{(\omega_0 t)},\nonumber\\
\langle x_{A}(\tau + t) x_{B}(\tau)\rangle &=& 
\frac{2(1+3\mu)}{3N}\; \frac{e^{-|\lambda|t}}{|\lambda|} \left\{\sqrt{3}\sin{(\omega_0 t)}-\cos{(\omega_0 t)}\right\}.
\end{eqnarray}
Following  Ref.~(\cite{Nisbet}), fluctuations are non-cyclic if the auto-correlation functions decay monotonically to zero, whereas they lead to quasi-cycles if the auto-correlations oscillate. In the latter case, quasi-cycles are ``phase-forgetting'' if the oscillations of the auto-correlation functions are damped, as in Eqs.~(\ref{autocorr}),  and ``phase-remembering'' when their amplitude do not vanish. Following this classification, it is clear from (\ref{autocorr})  and  Figs.~\ref{StochRes} and \ref{StochRes_regb} that the quasi-cycles in the region (a) of the parameter space are ``phase-forgetting'', while those in  the region (b) are ``phase-remembering''.

An analysis similar to that of this section can be carried out in the region (b) of the phase parameter by performing a van Kampen expansion around the limit cycle $\bar{\bm \sigma}(t)$. This leads to deal with a stochastic version of 
non-autonomous linear differential equations~(see, e.g., (\cite{Boland})).
\subsection{Average escape time}
\begin{figure} 
\begin{center} 
\includegraphics[scale=0.5]{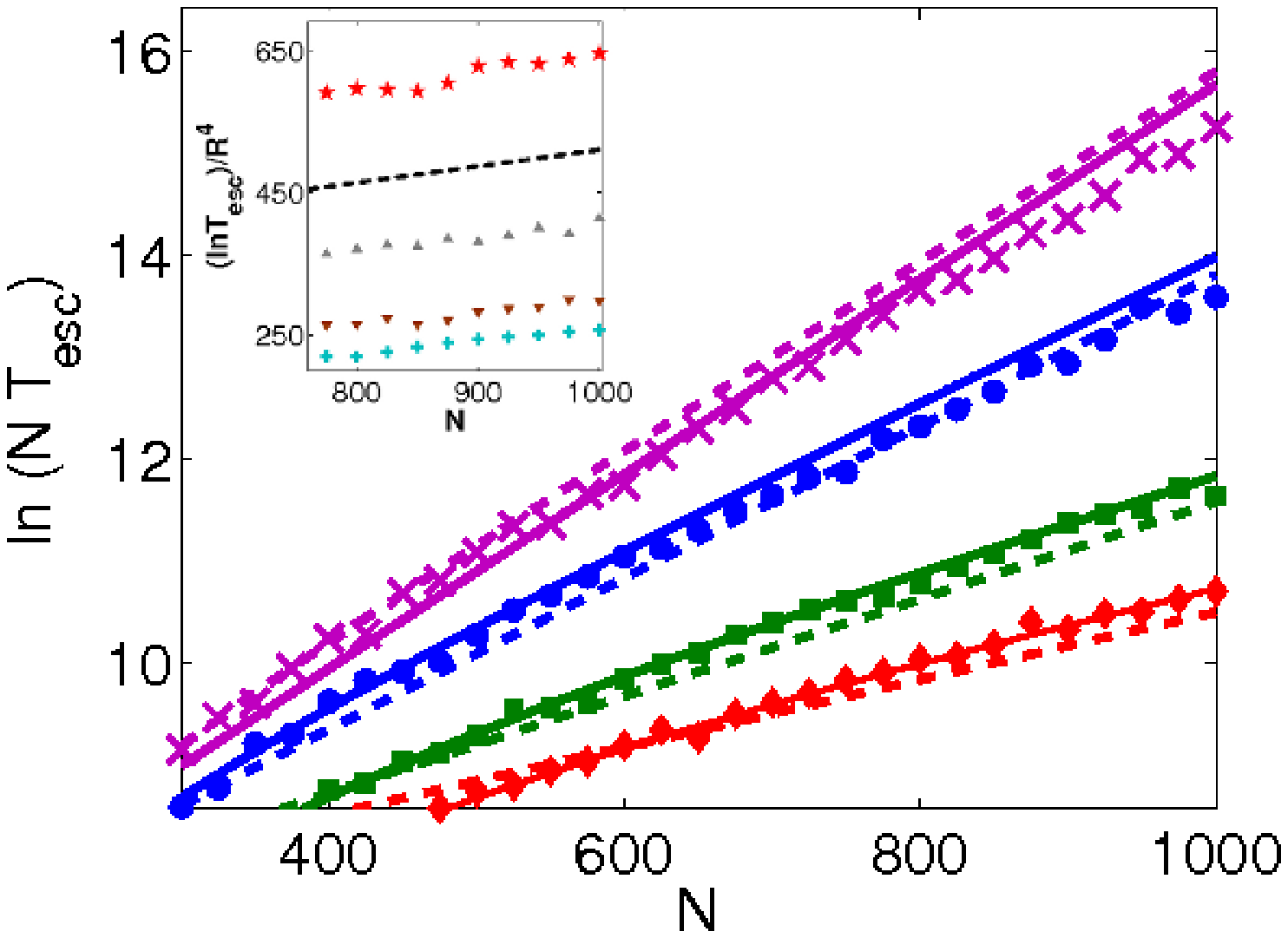}
\includegraphics[scale=0.355]{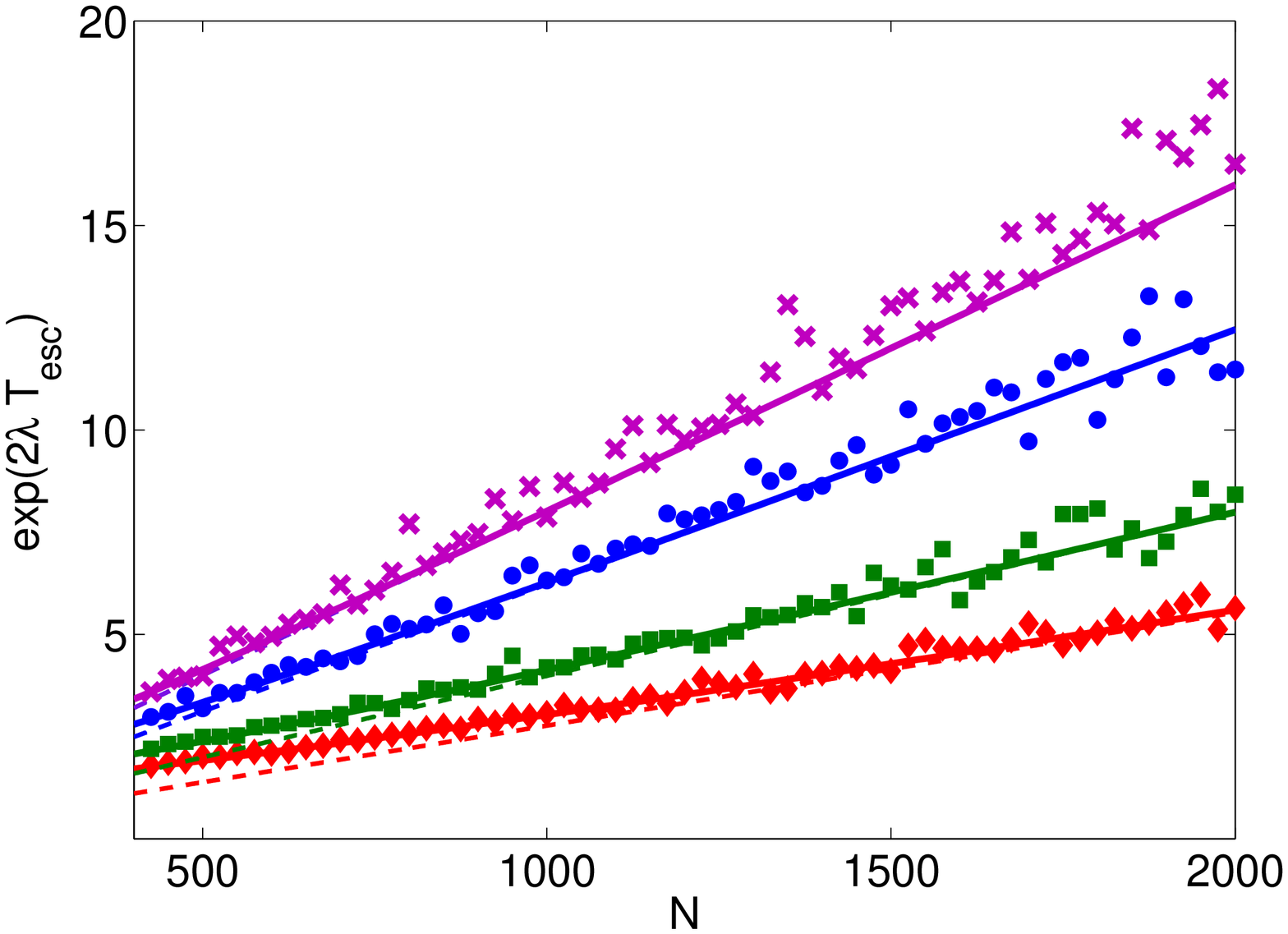}
\caption{{\it (Color online)}. Average escape time $T_{\rm esc}(R)$ to reach the cycles ${\cal C}(R)$ starting from  ${\bm s^*}$, for $R=0.10~(\Diamond), 0.12~(\Box), 0.15~(\bullet), 0.17~(\times)$ and large population size $N$.
Left: Plot of  $\ln{(N T_{\rm esc}(R))}$ versus $N$ in the case $\lambda<0$, with $\epsilon=0.5$ and $\mu=0.01$, and sample-average over $100$
 replicates. We find a linear dependence that yields $T_{\rm esc}(R)\propto e^{C_1(R)~N}/N$. Numerical results (filled  symbols) are also compared to the predictions of Eqs.~(\ref{Tex}) (solid) and (\ref{Tex-1}) (dashed), yielding $C_1(R)\propto R^2$.
Right: Plot of ${\rm exp}{(2\lambda T_{\rm esc}(R))}$ versus $N$ in the case $\lambda>0$, with $\epsilon=1.5$ and $\mu=0.01$ and sample-average over $400$ replicates. The observed linear dependence 
 yields  $T_{\rm esc}(R)\propto \ln{(C_2(R)~ N)}/(2\lambda)$. 
Numerical results (filled  symbols) are also compared to the predictions of Eqs.~(\ref{Tex}) (solid) and (\ref{Tex-2}) (dashed lines, almost indistinguishable from the solid lines), yielding $C_2(R)\propto R^2$.
 Inset: Marginal case $\lambda=0$ with parameters $\epsilon=1.18$ and  $\mu=0.01$, and sample-average over $100$ replicates. Plot of $R^{-4}\ln{(T_{\rm esc}(R))}$ versus $N$ for values of $R=0.30~(\star), 0.35~(\triangle),0.40~(\nabla),0.45~(+)$. For large population size, the (approximate) linear dependence $R^{-4}\ln{(T_{\rm esc}(R))} \propto 0.23 N$ (dashed line as a guide for the eyes) yields $T_{\rm esc}(R)\sim e^{0.23R^4N}$  (see text).
\label{TescPlot}}
\end{center}
\end{figure}

Another intriguing effect of demographic noise is related to the pivotal concepts of {\it attractor} and {\it stability}. As discussed in the previous sections there are two different types of stable attractors in the model RPSM: ${\bm s}^{*}=(1/3,1/3,1/3)$ is the only  attractor of the system when $\lambda\leq 0$, whereas  all trajectories in the phase portrait approach the limit cycle $\bar{{\bm \sigma}}=(\bar{a}(t),\bar{b}(t))$ when $\lambda> 0$.  This scenario predicting drastically different fate depending on whether  
$\lambda\leq 0$ or $\lambda>0$ has to be revised when fluctuations are accounted.
 In this case, within a probabilistic setting, 
attractors can be regarded 
  as minima of a potential well from which it is   {\it always} possible to escape  after some characteristic time (which can be enormously long, see, e.g., (\cite{Gardiner,Risken,Kubo,Dykman,Volovik})). Stochasticity affects the concept of stability
even if the system is initially at the fixed point ${\bm s}^{*}$. In fact, due to demographic fluctuations, the species densities deviate from ${\bm s}^{*}$ and yield quasi-cycles resulting in persistent erratic oscillations of non-vanishing amplitude. It is therefore biologically relevant to assess the robustness of the population composition and understand how noise affects the co-evolution of sub-populations that initially coexist  with the same density. 
Here, to further assess the influence of intrinsic fluctuations when $N$ is large yet finite, we are interested in the average time to escape from  ${\bm s}^{*}$ and reach a specified separating distance from it. In other words, we compute the average time $T_{{\rm esc}}(R)$ for a trajectory starting at  ${\bm s}^{*}$ to reach a cycle ${\cal C}(R)$ on which oscillations around ${\bm s}^{*}$ are of amplitude $2R/\sqrt{3}$ (see below). Below, we discuss our results, that are summarized in  Fig.~\ref{TescPlot}, and also present an analytical approach -- based on the diffusion theory and van Kampen expansion -- to compute ${\cal T}_{{\rm esc}}(R)$, that is an accurate approximation of $T_{{\rm esc}}(R)$ around  ${\bm s}^{*}$. 

Using the Gillespie algorithm, we have computed numerically the average escape time $T_{{\rm esc}}(R)$ necessary to reach a cycle ${\cal C}(R)$ starting from the interior fixed point $a^*=b^*=c^*=1/3$ (same initial density of each species). To exploit the symmetry of the model (see Sec. 3.2), we work with the variables ${\bm y}=(y_A,y_B)={\cal S}{\bm x}$ and 
consider cycles ${\cal C}(R)$ of parametric equation $y_A=R\cos{\phi},\; y_B=R\sin{\phi}$. In the original ${\bm x}-$variables, the equation of ${\cal C}(R)$ is $x_A=R\left(\cos{\phi}-\sin{\phi}/\sqrt{3}\right),\, x_B=(2R/\sqrt{3})\sin{\phi}$ and the amplitude of the oscillations is $2R/\sqrt{3}$. 
In our numerical computations, we have considered systems of population size $N$, with $N=300 - 4000$, and values of $R$ ranging from $0.10$ to $0.45$. Each simulation has been sample-averaged over $100$ to $400$ replicates. As we are mainly interested in the linear regime around ${\bm s}^{*}$, where analytical calculations are amenable (see below), particular attention has been dedicated to values of $R\approx 0.10 - 0.20$ (i.e. amplitude $\approx 0.11 - 0.23$). Typical results are reported in Fig.~\ref{TescPlot}, where one essentially distinguishes two cases. If $\lambda<0$,  ${\bm s}^*$ is stable and  $\ln{(N T_{\rm esc}(R))}$ grows linearly with $N$ (for large $N$) with a slope depending on $R$,  as illustrated in Fig.~\ref{TescPlot}~(left), which yields $T_{\rm esc}(R)\propto e^{C_1(R)N}/N$. On the other hand, when  ${\bm s}^*$ is unstable ($\lambda>0$)
and $N$ is large, we find that ${\rm exp}(2\lambda T_{\rm esc}(R))$ varies linearly with $N$ (slope depending on $R$), as shown in Fig.~\ref{TescPlot}~(right). This leads to  $T_{\rm esc}(R)\propto \ln{(C_2(R)~ N)}/(2\lambda)$. In the above asymptotic expressions, $C_1(R)$ and $C_2(R)$ are monotonic functions of $R$ and comparison with analytical calculations around ${\bm s}^*$ suggests that $C_1 \sim C_2 \propto R^2$ (see below). For the marginal case $\lambda=0$ and  large $N$,
numerical results are reported in  the inset of Fig.~\ref{TescPlot}~(left) and we find that $(\ln{T_{\rm esc}(R))/R^4}$ (approximately) exhibits a linear dependence on $N$ (with constant slope). In the case $\lambda=0$, one thus finds  $T_{\rm esc}(R)\sim e^{C_3 R^{4} N}$, where $C_3$ is a constant.

 In the realm of the diffusion theory and van Kampen expansion, the calculation of the (approximate) average escape time ${\cal T}_{{\rm esc}}({\bm x})$ can be regarded as a first-passage problem 
 to an absorbing cycle ${\cal C}(R)$ starting from a system initially at ${\bm s}^*$~(\cite{Gardiner,Redner}). To obtain an equation for  ${\cal T}_{{\rm esc}}$ it is useful to consider the so-called backward Kolmogorov equation (BKE), which is 
 the adjoint of Eq.~(\ref{FPE}) and reads (see, e.g., (\cite{Gardiner,Risken,Redner})):
\begin{eqnarray}
\label{Lb}
-\partial_t P_{b}({\bm x},t)&=&{\cal L}_{b} P_{b}({\bm x},t), \nonumber\\  \text{with} \quad
{\cal L}_{b}({\bm x})&=& {\cal A}_{ij}({\bm s}^{*}) x_j \; \partial_{x_i}  +\frac{1}{2}\,{\cal B}_{ij}({\bm s}^{*})\partial_{x_i} \partial_{x_j}.
\end{eqnarray}
As in Section 3.2, to reveal the polar properties of the system in the vicinity of ${\bm s}^{*}$, it is natural to perform the linear transformation ${\bm x}\to {\bm y}={\cal S}{\bm x}$ and work in the ${\bm y}-$variables.
Under this transformation,
 one has $P_{b}({\bm x},t) \to {\widetilde P}_{b}({\bm y},t)$, ${\cal A}({\bm s}^{*}) \to {\cal S}{\cal A}({\bm s}^{*}){\cal S}^{-1}=\begin{pmatrix}
\lambda & -\omega_0 \\
\omega_0 & \lambda
\end{pmatrix}$ and ${\cal B}({\bm s}^{*}) \to {\cal S}{\cal B}({\bm s}^{*}){\cal S}^{T}=\frac{1+3\mu}{3N}~\begin{pmatrix}
1 & 0 \\
0 & 1
\end{pmatrix}$~(\cite{reichenbach-2006-74}). Thus, in the ${\bm y}-$variables,  the differential operator of Eq.~(\ref{Lb}) becomes ${\cal L}_{b}({\bm x}) \to {\widetilde
{\cal L}}_{b}({\bm y})= \lambda(y_A \partial_{y_A}+y_B \partial_{y_B}) + \omega_0(y_B \partial_{y_A}-y_A \partial_{y_B})-\frac{1+3\mu}{6N}(\partial_{y_A}^2+\partial_{y_B}^2)$.
To exploit the system's symmetry around ${\bm s}^{*}$, we then adopt
the polar coordinates $(\rho,\phi)$, with $y_A=\rho \cos{\phi}$ and $y_B=\rho \sin{\phi}$. 
The radial BKE thus reads $\partial_t {\widetilde P}_{b}(\rho,t)={\widetilde {\cal L}}_{b}(\rho) {\widetilde P}_{b}(\rho,t)$, with
\begin{eqnarray}
\label{Lb-radial}
{\widetilde {\cal L}}_{b}(\rho)=\left[\lambda \rho +\left(\frac{1+3\mu}{6N}\right) \frac{1}{\rho}\right]\partial_{\rho} + \left(\frac{1+3\mu}{6N}\right)\partial_{\rho}^{2}
\end{eqnarray}
 To derive the above radial operator,
we have assumed that the initial radial symmetry of the probability distribution (initially the system is at ${\bm s}^{*}$, without any angular dependence) is approximately preserved by the dynamics. The above approach is certainly valid at linear order around ${\bm s}^{*}$ and is consistent with the van Kampen linear expansion that has led to (\ref{FPE}) and (\ref{Lb}).
With (\ref{Lb-radial}), the equation for the average escape time ${\cal T}_{{\rm esc}}(\rho)$ at a distance $\rho$ from ${\bm s}^{*}$ is given by~(\cite{Gardiner,Risken}) 
\begin{eqnarray}
\label{Tex-eq}
-1= {\widetilde {\cal L}}_{b}(\rho)~{\cal T}_{{\rm esc}}(\rho).
\end{eqnarray}
In the framework of diffusion and van Kampen approximations, 
we now analytically compute  the  mean escape time ${\cal T}_{{\rm esc}}(R)$ for
$ \rho$ to attain 
the value $R$ starting from ${\bm s}^{*}$ (i.e. initially $\rho=0$), which corresponds to the mean time necessary to reach the cycle
${\cal C}(R)$.
Our analytical treatment is valid in the linear regime around ${\bm s}^{*}$, where 
$\phi\simeq \omega_0$, and the amplitude of the oscillations on ${\cal C}(R)$ is $2R/\sqrt{3}$.
In our theoretical setting, the average escape time ${\cal T}_{{\rm esc}}(R)$ is obtained by solving
 Eq.~(\ref{Tex-eq})  subject to 
a reflecting and an absorbing boundary conditions at $\rho=0$ and $\rho=R$, respectively~(\cite{Gardiner}).
When $\lambda \neq 0$, the solution to this problem can be expressed in terms of the function $\Psi(z)=z\,{\rm exp}\left(\frac{3N\lambda}{1+3\mu}~z^2\right)$:
\begin{eqnarray}
\label{Tex}
{\cal T}_{{\rm esc}}(R)&=& \left(\frac{6N}{1+3\mu}\right)~\int_{0}^{R}\frac{dy}{\Psi(y)}~\int_{0}^{y}~dz\Psi(z)=\frac{1}{2\lambda}\int_{0}^{-\frac{3NR^2\lambda}{1+3\mu}} \frac{du}{u}~ \left(1-e^u\right).
\end{eqnarray}
The solid curves in Fig.~\ref{TescPlot} show the comparison between 
the prediction (\ref{Tex}) and numerical results. We notice that for small values of $R$, e.g., $R=0.10 - 0.15$, there is an excellent agreement between (\ref{Tex}) and the results of numerical simulations (both in the cases $\lambda<0$ and $\lambda>0$). For larger values of $R$ (e.g. for $R\geq 0.17$), nonlinearities and non-vanishing angular dependence cause some systematic deviations from the results of numerical computations (discrepancies of  $\approx 5\% - 10\%$ in the results of Fig.~\ref{TescPlot} for $R=0.17$), but (\ref{Tex}) still provides the correct (qualitative) behavior and a reasonable approximation
of $ T_{{\rm esc}}(R)$.
In the asymptotic limit of large population size $N$,
two distinct behaviors can be obtained from Eq.~(\ref{Tex}):
\begin{itemize}
 \item When $\lambda<0$,  ${\bm s}^{*}$ is stable and
the main contribution to  (\ref{Tex}) is given by
${\cal T}_{{\rm esc}}(R)\simeq \frac{1}{2|\lambda|}~{\rm Ei}\left(\frac{3|\lambda|R^2 N}{1+3\mu}\right)$, where ${\rm Ei}(x)\equiv \int_{-\infty}^{x} dt\,(e^{t}/t)$ is the exponential integral~(\cite{Abramowitz}). From the properties of this function, we infer the following asymptotic behavior (for $|\lambda|NR^{2} \gg 1$):
\begin{eqnarray}
\label{Tex-1}
{\cal T}_{{\rm esc}}(R)\simeq \left(\frac{1+3\mu}{6(\lambda R)^2 N}\right)\,{\rm exp}\left(
\frac{3|\lambda|R^2 N}{1+3\mu}
\right)
\end{eqnarray}
This result predicts that the mean escape time increases dramatically with the total number of individuals $N$ and, when
$|\lambda| NR^{2} \gg 1$, is  proportional to the exponential of the population size. The dashed lines in Fig.~\ref{TescPlot} (left) illustrate that  (\ref{Tex-1})  is a very  good approximation of $T_{{\rm esc}}(R)$ in  the linear regime around ${\bm s}^*$, e.g. for $R=0.10 - 0.15$, when $N\gg 1$~\footnote{In Fig.~\ref{TescPlot} (left), $3|\lambda|R^2 N /(1+3\mu)$ lies between $\approx 1.0$ (for $R=0.1$ and $N=300$) and $\approx 9.5$ (for $R=0.17$ and $N=1000$).}. In particular, we notice that (\ref{Tex-1}) coincides with
the aforementioned asymptotic expression of $T_{{\rm esc}}(R)$  with $C_1=3|\lambda| R^2 / (1+3\mu)$. For $R=0.17$, when  $N$ is large, nonlinear effects are important and in  Fig.~\ref{TescPlot} the exponent of (\ref{Tex-1}) overestimates that of $T_{{\rm esc}}(R)$ by  $\approx 5\%$.
\item
When $\lambda>0$,  
demographic fluctuations always cause the departure from ${\bm s}^{*}$ (unstable), and 
all trajectories in the phase portrait are attracted by the limit cycle $\bar{{\bm \sigma}}$. In this case, from the properties of the  exponential integral~(\cite{Abramowitz}), for $\lambda NR^{2} \gg 1$, the main contribution to (\ref{Tex}) is
\begin{eqnarray}
\label{Tex-2}
{\cal T}_{{\rm esc}}(R)\simeq \frac{1}{2\lambda}~\left[\ln{\left(\frac{3\lambda R^2 N}{1+3\mu}\right)}+\gamma_{\rm EM} \right],
\end{eqnarray}
where $\gamma_{\rm EM}= 0.57721...$ is Euler-Mascheroni's constant.
This result predicts that, when $\lambda NR^{2} \gg 1$, the mean escape time 
grows logarithmically with the population size $N$, i.e. ${\cal T}_{{\rm esc}}(R)\to (\ln{N})/2\lambda$, and the value of $R$ contributes to subdominant corrections.  Here also, 
the dashed lines in Fig.~\ref{TescPlot} (right), almost indistinguishable from solid curves, agree very well with 
the numerically computed $T_{{\rm esc}}(R)$ around ${\bm s}^*$ (i.e. for $R=0.10 - 0.15$).
We notice that, when $C_2=3\lambda R^2 / (1+3\mu)$, the prediction (\ref{Tex-2}) coincides with
the asymptotic expression of $T_{{\rm esc}}(R)$ inferred from Fig.~\ref{TescPlot} (right).
\end{itemize}

It is also worth noticing that results (\ref{Tex})-(\ref{Tex-2}) are also valid in the absence of mutations (i.e. when $\mu=0$).

Besides the above cases, a special situation arises when $\lambda=0$, $\mu>0$ and ${\bm s}^{*}$  is a center. In fact, in Sec. 3.2, we have seen that in this case ${\bm s}^{*}$ is an attractor, but  the dynamics in its vicinity is very slow. To account for such a slow dynamics it is not sufficient to consider (\ref{FPE}, \ref{Lb}) -- obtained by linearization about ${\bm s}^{*}$ -- but nonlinearities have to be accounted. Generalizing the above approach and keeping the cubic terms in the drift contribution (\cite{Cremer}), leads to (\ref{Tex-eq}) with the following differential operator  [instead of (\ref{Lb-radial})]: $
{\widetilde {\cal L}}_{b}(\rho)=\left[\{\lambda+\beta \rho^2\} \rho +\left(\frac{1+3\mu}{6N}\right) \frac{1}{\rho}\right]\partial_{\rho} + \left(\frac{1+3\mu}{6N}\right)\partial_{\rho}^{2}$. When $\lambda=0$,  proceeding as above, one still finds an exponential asymptotic behavior ($\mu R^{4} N \gg 1$): ${\cal T}_{{\rm esc}}(R)\sim N^{-1/2} \, {\rm exp}(27\mu R^4 N/(1+3\mu))$. 
This estimate, that is consistent with the numerical results reported in Fig.~\ref{TescPlot}~(left, inset), implies that (at given $R$ and $N$) the mean escape time ${\cal T}_{{\rm esc}}$ around ${\bm s}^{*}$
is significantly shorter in the marginal case  than in the case  $\lambda <0$.

\section{Summary and Conclusion}

This work has been motivated by the  importance of further understanding the mechanisms at the origin of persistent and erratic oscillatory dynamics in population biology, which is frequently observed but whose theoretical origin is often 
debated. Furthermore, this work concerns the effects of mutations in rock-paper-scissors (RPS) games, which are regarded as paradigmatic models to describe the co-evolutionary dynamics of systems with codominant interactions. It has recently been proposed that codominance can help  promote the maintenance of biodiversity, and it is therefore 
of biological relevance to study generalizations of the RPS model to understand which ingredients  favor the long-term coexistence of the sub-populations.

Here, we have 
investigated the oscillatory dynamics of the generic rock-paper-scissors games with mutations in a well-mixed (homogeneous) population of $N$ individuals. Our study has been carried out in the mean-field limit ($N\to \infty$) and in the presence of demographic  noise ($N$ large but finite).
In addition to the regions of the parameter space associated with a stable interior fixed point ${\bm s}^*$ and heteroclinic cycles, the possibility for the individuals to switch from one strategy (species) to another with a small transition rate yields
a new oscillatory behavior associated with a limit cycle and resulting from a Hopf bifurcation.
In the mean-field limit ($N\to \infty$), we have recast the system's rate equations in a normal form and 
studied the main properties of the ensuing limit cycle. When the population size is finite, demographic noise has to be taken into account and the dynamics becomes genuinely stochastic {\it and} nonlinear.
The latter situation has been described in terms of an individual-based formulation,
with the stochastic dynamics
implemented  according to a Moran process. The influence of intrinsic noise on the dynamics has been analytically assessed within a diffusion theory approximation complemented by numerical simulations (using the Gillespie algorithm). We have shown that demographic noise transforms damped oscillations into (``phase-forgetting'') quasi-cycles and perturbs the amplitude and the phase of limit cycles (that become ``phase-remembering'' quasi-cycles). As also observed in other systems~(e.g., (\cite{Bartlett1,Bartlett2,Nisbet,McKane1,Boland})), the effect of stochasticity is particularly striking in the region  of the parameter space where ${\bm s}^*$ is stable. There, we have found sustained oscillations driven by demographic stochasticity. In fact, while fluctuations are of order $N^{-1/2}$, for small mutation rate, the amplitude of the oscillations is amplified by a resonance amplification caused by internal (white) noise. We have shown that this phenomenon translates into an isolated and well-marked peak in the power-spectrum and have also computed the autocorrelation functions of the system.
To further assess the robustness of the co-evolutionary dynamics against noise, we have computed numerically and analytically -- using a diffusion approximation and a mapping onto a first-passage problem -- the average escape time necessary to reach a cycle on which the oscillations attain a given amplitude. We have therefore shown that such a mean exit time grows either exponentially or logarithmically with the system size, depending on whether the interior fixed point is deterministically stable or unstable, respectively.

In summary, 
in the presence of mutations the RPS model yields sustained oscillations in every region of the parameter space. The latter are caused by demographic noise and/or result from a Hopf bifurcation. 
Therefore, mutations in the RPS model ensure that all
species co-evolve and oscillate in time without ever going extinct. 
One can therefore regard  mutations as a mechanism favoring the oscillatory  dynamics in communities characterized by cyclic co-dominance.

In future research, it would be interesting to consider spatially-extended versions of this model and study how the population will self-organize and which kinds of spatial-temporal patterns would emerge. In particular,
as it has recently been shown that mobility can affect co-evolution in RPS models by mediating between mean-field and stochastic dynamics~(\cite{reichenbach-2007-448,reichenbach-2007-99,reichenbach-2008}), it would be  relevant to investigate variants of the model RPSM where individuals are allowed to move and interact in space.

\section*{Acknowledgments}
This work was initiated with the support of the Swiss National Science Foundation through the Advanced Fellowship PA002-119487.

{\small

\end{document}